\documentclass{aa}  

\usepackage{graphicx}

\usepackage{txfonts}
\usepackage{graphicx}   
\usepackage{amsmath}    
\usepackage{amssymb}    
\usepackage{multicol}        
\usepackage{bm}         
\usepackage{color}
\usepackage{xcolor}
\usepackage{soul}         
\usepackage{tabularx}

\newcommand\clearrow{\global\let\rowmac\relax}
\clearrow

\begin{document} 
        
        \titlerunning{New {\em TESS} pulsating subdwarfs}
        \title{Discovery of new {\em TESS} pulsating hot subdwarfs}
        
        \author{ J.~Krzesinski\inst{1}\and L.~A.~Balona\inst{2}}
        
        \institute{Astronomical Observatory, Jagiellonian University, 
                ul. Orla 171, PL-30-244 Krakow, Poland
                \\
                \email{jk@oa.uj.edu.pl}
                \and
                South African Astronomical Observatory, P.O. Box 9, Observatory, Cape
                Town, South Africa
                \\
        \email{lab\@saao.ac.za}
        }
        \date{Received 2021}
        
        
        \abstract
        {}
        {
        This work is dedicated to a search for new pulsating hot 
        subdwarfs in {\em TESS} photometric data which could have been missed 
        in previous searches.           
        }
        {
        By matching catalogues of hot subdwarfs with {\em TESS} targets and using 
        luminosities from {\em Gaia} parallaxes, a list of 1389 candidate hot subdwarfs 
        observed by {\em TESS} was created. The periodograms of these stars were 
        inspected, and the stars were classified according to variability type.                    
        }
        {
        An updated catalogue of all known pulsating hot subdwarfs is
        presented. A number of probable pulsating binaries have been 
        identified, which might prove useful for verifying the asteroseismic 
        masses.  The mean masses of p- and g-mode pulsators are  estimated 
        from the stellar parameters.
        }
        {
       A list of 63 previously unknown pulsating hot subdwarfs observed by {\em
        TESS} is presented.  More than half of the stars previously identified 
        as pure p-mode pulsators are found to have frequencies in the g-mode 
        region as well.  As a result, hybrid p- and g-mode pulsators occur 
        over the whole instability strip.       
        }
        
        
        \maketitle

\section{Introduction}

Hot subdwarf stars form roughly two classes of objects. The cooler subdwarf B-type (sdB) class, whose spectra typically show no or weak helium lines, and the hotter subdwarf O-type (sdO) class, which has (on average) a higher helium abundance \citep{Stroeer2007}. 

The sdBs are core He-burning stars on the extreme horizontal branch having very thin hydrogen envelopes \citep{Heber1986, Saffer1994,Heber2016}. It is thought that most of the hydrogen was removed by 
mass loss that could have occurred as a result of mass interchange in a binary
system \citep{Han2002,Han2003}. This is supported by the fact that many 
sdB stars are close binary systems, which are either 
late M-type main sequence stars or white dwarfs \citep{Maxted2001,Heber2016}. 
The effective temperature range of the sdBs is between 20\,000 and 40\,000\,K and the surface gravity ($log\,g$) is between 5.2 and 6.2 \citep{Green2008,Saffer1994}.       
        
The hotter sdO class encompasses a wider range of hot objects with diverse origins. These stars are located blueward of the extreme horizontal branch stars in the HR diagram \citep{Stroeer2007,Nemeth2012}. Most are likely descendants of red giant stars; however, a few have luminosities and temperatures similar to the central stars of planetary nebulae and therefore might be of post-AGB origin \citep{Heber1988, Rauch1991}. Because evolutionary tracks of sdB stars pass through the sdO region \citep{Dorman1993}, a link between the two classes is also possible and some sdOs in the class might  also be descendants of sdB stars. On the other hand, the population of   sdO stars consists mostly of single stars \citep{Napiwotzki2008}; therefore, a merger of two helium white dwarfs  \citep{Webbink1984, Iben1984,Saio2000}, and the delayed helium flash of a WD \citep{Lanz2004} or hot-flasher \citep{Miller2008}  were the scenarios suggested to explain the existence of some of these stars.
        
The effective temperatures of the hotter sdOs can reach up to 80\,000\,K, and due to the diversity of the sdO stars, their surface gravity ($log\,g$) range (4.0 -- 6.5)  is broader than for sdBs \citep{Oreiro2004,Johnson2014,Heber2009, Heber2016}. The fraction of sdO stars with temperatures around 40\,000\,K are likely He-burning subdwarfs, while the hotter sdOs are thought to have entered a subsequent He-shell-burning phase \citep{Wang2021}. For an in-depth review of the two classes, see \citet{Heber2016}.

\citet{Charpinet1996} investigated pulsational driving in sdB stars. They 
found that most of the driving is the result of the opacity $\kappa$-mechanism 
operating in the partial ionisation zone of Fe-group elements.  This is the
same mechanism that operates in the main sequence $\beta$~Cep and SPB 
stars \citep{Moskalik1992, Dziembowski1993b}.  However, pulsation in models
of sdB stars will only occur if the metal abundance in the driving region 
is enhanced by about a factor of two at uniform envelope
abundance. A basic result of radiatively driven diffusion is that 
an element accumulates in a layer where the specific opacity for the element is
highest. Therefore, Fe-group elements will tend to be enhanced in the region of
the opacity bump, as required for pulsational driving.

Shortly after these model calculations, \citet{Kilkenny1997} reported the 
discovery of small-amplitude, rapid light variations in the sdB binary, 
EC\,14026-2647.  The main frequency peak is at about 600\,d$^{-1}$ with an 
amplitude of about 0.01\,mag.  Since then, many other sdB stars pulsating at
high frequencies have been discovered. These p-mode pulsators are now called 
the V361~Hya variables.

\citet{Green2003} discovered a second family of related sdB stars.  These
are also low-amplitude multiperiodic variables, but with frequencies in the 
range 10--50\,d$^{-1}$. These are called V1093~Her variables.  The
pulsations are attributed to g-modes of low spherical harmonic degree driven 
by the same mechanism as in the hotter V361~Hya stars \citep{Fontaine2003}.  
Because they have lower effective temperatures, the driving zone is deeper in 
the V1093~Her stars and the thermal timescale is longer.  As a result, only 
g-mode pulsations of long periods are unstable. The situation is analogous to 
the $\beta$~Cep/SPB case.

It should be mentioned that pulsating stars have also been  found among sdOs. The first in this class is SDSS J160043.6+074802.9, a subdwarf discovered by \citet{Woudt2007}; it is a binary with a late-type companion \citep{Rodriguez2010}. The second,  previously classified as a sdBV in a binary system with an F0 star \citep{Koen1997,ODonoghue1998}, is EO\,Cet (PB8783), which was later re-classified as an sdO pulsator \citep{Ostensen2012}. The last is  EC\,03089-6421, a sdOV star discovered by \citet{Kilkenny2017}. All are rapid pulsators with periods in the range 0.5 -- 2 minutes. There were also cooler ($\sim$50\,000\,K) sdO pulsators discovered in a globular cluster ($\omega$ Cen) \citep{Randall2011,Randall2016}, but they likely make up a different class of subdwarf pulsators than the field sdOV stars.

Space photometry from the {\em Kepler} mission \citep{Borucki2010} and from
the {\em TESS} mission \citep{Stassun2019} has been an important source of new 
candidates for V1093~Her and V361~Hya
classes of variables \citep{Ostensen2010b, 
Ostensen2011c, Kawaler2010a, Kawaler2010b, Reed2010}.
The majority of the sdB stars observed by {\em Kepler} are V1093~Her
variables (16 stars), with one hybrid and one V361~Hya star.  
Analysis of data from the extended {\em Kepler K2} mission resulted in the 
discovery of 15 pulsating sdB variables, two of which belong to the V361~Hya class \citep{Reed2018}.  Analysis of data from 
the {\em TESS} mission is ongoing \citep{Reed2020, Uzundag2021, Sahoo2020, Charpinet2019}.  
\citet{Holdsworth2017} gives a table of 110 pulsating subdwarfs known at the time.  \citet{Lynas-Gray2021}
presents a good review which includes a list of 56 known pulsating sdB stars.

More recently, \citet{Groot2021} has presented three lists of 1302, 269, and 1013 
confirmed hot subdwarfs (including hot subdwarf pulsators) observed by the
{\em TESS} mission. These numbers are taken from their online catalogues (see 
link in their paper). These catalogues were assembled by Working Group 8 (WG8, 
compact pulsators) of the {\em TESS Asteroseismic Consortium}.  Their contents
overlap, so that the total number of entries is larger than the actual number 
of hot subdwarfs.

In this paper we present the results of a search for pulsating hot subdwarfs
using {\em TESS} photometry.  We begin in Section 2 by describing the 
{\em TESS} data and how hot subdwarfs are identified.  We emphasize  the 
importance of {\em Gaia} parallaxes \citep{Gaia2016, Gaia2018} in this
regard. In Section 3 we discuss lists of known hot subdwarf pulsators and why 
it is important to identify p- and g-mode pulsators. 
In Section 4 the search for variability among 
{\em TESS} stars identified as probable hot subdwarfs is discussed.  The 
criteria used to distinguish between eclipsing and pulsating stars from the 
periodograms is described.  The effect of contamination from nearby stars is
discussed.
The mass distribution of hot subdwarfs derived from {\em Gaia} luminosities and 
published surface gravities is discussed in Section 5. 
This allows the relative masses of p- and g-mode pulsating hot subdwarfs to be  estimated.
In Section 6 the location of the pulsating hot subdwarfs in the Hertzsprung--Russell (H--R) diagram
is discussed and compared to the models. In the Appendix a catalogue of all 
known pulsating hot subdwarfs discovered since the previous listing by 
\citet{Holdsworth2017} is presented.  A table of the newly discovered  
pulsating hot subdwarfs is also presented.  A separate list of many new 
pulsating hot subdwarfs, which were unrecognized as 
pulsators by \citet{Groot2021}, is included.
Finally, the periodograms of all newly discovered pulsating hot subdwarfs
are shown.

\section{Data and identification of hot subdwarfs}

The {\em TESS} photometric survey consists of continuous wide-band photometry 
of 13 sectors per celestial hemisphere \citep{Ricker2014}.  Each sector is 
observed continuously for about 27\,days with 2 min cadence.  Stars near the 
ecliptic equator are only observed for one sector. Stars nearer 
the poles may be observed in more than one sector so that, at the ecliptic poles, 
stars are observed continuously for over 350\,days. The light curves are 
corrected for time-correlated instrumental signatures using pre-search data 
conditioning (PDC, \citealt{Jenkins2016}). These are the data labelled  
{\tt PDCSAP\_FLUX} in the light curve FITS files.  The data used here are the full 
\mbox{\tt PDCSAP} light curves from sectors 1--38.  As this was part of a
survey for stellar variability among many thousands of {\em TESS} stars, no 
effort was made to optimise the aperture.  The light curves for {\em TESS} 
stars are available on the {\em MAST} website 
({\tt https://archive.stsci.edu/}).

Hot subdwarfs were first identified in surveys of faint blue stars at high 
Galactic latitudes \citep{Humason1947}. Subsequent surveys added to the
list.  The first catalogue of spectroscopically identified hot subdwarf stars
\citep{Kilkenny1988} contained 1225 sdB and sdO stars. \citet{Ostensen2006} 
expanded the list to more than 2300 stars.  The advent of mass surveys such
as the Sloan Digital Sky Survey (SDSS) extended the search to fainter
magnitudes and provided spectra of almost 2000 hot subdwarfs
\citep{Geier2015, Kepler2015, Kepler2016}.  The precise proper motions
and parallaxes provided by {\em Gaia} \citep{Gaia2016, Gaia2018} has led to an 
even greater number of possible subdwarf candidates based on colour, absolute 
magnitude, and reduced proper motion cuts \citep{Geier2017, Geier2019, 
Geier2020}.  

These lists were cross-matched with the {\em TESS} input catalogue
\citep{Stassun2019}, leading to 1398 hot subdwarf candidates for 
which light curves are available. Not surprisingly, the majority of these 
stars are also listed on the WG8 list of compact pulsators.

\begin{figure}
\centering
\includegraphics[]{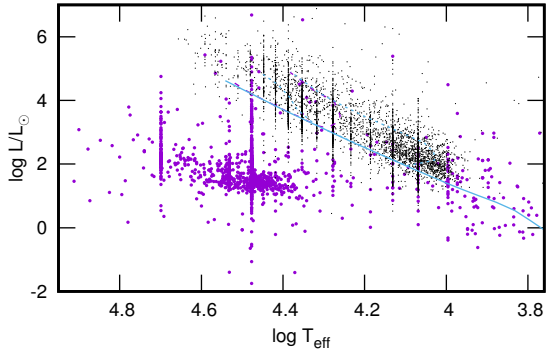}
\caption{Theoretical H--R diagram showing subdwarfs (violet filled
circles) and main sequence stars brighter than 12.5\,mag (small black circles) 
observed by {\em TESS}. Luminosities were derived from {\em Gaia} parallaxes.  
The vertical arraignment of some points is due to adopted values of $T_{\rm eff}
= 50000$\,K for sdO and $T_{\rm eff} = 30000$\,K for sdB stars for which values
are not available and the use of $T_{\rm eff}$ from spectral types for some
main sequence stars.  The solid line is the zero age main sequence for models 
with solar abundance \citep{Bertelli2008}.  Also shown are the theoretical 
instability strips for $\beta$~Cep and SPB stars from \citet{Miglio2007}. 
}
\label{hrdiag}
\end{figure}

Perhaps the most direct way of identifying hot subdwarfs is to place the
star in the H--R diagram.  The photometric effective temperatures for hot 
stars are unreliable unless they include observations in the UV band.  The 
sources of effective temperature, $T_{\rm eff}$, are mainly those in 
\citet{Geier2020}, but the literature was searched for further values using 
the {\em SIMBAD} database  \citep{Wenger2000} for stars of interest.
The luminosity for each star was estimated from {\it Gaia} EDR3 
parallax  \citep{Gaia2016, Gaia2018} in conjunction with reddening obtained 
from a three-dimensional map by \citet{Gontcharov2017} using the bolometric 
correction calibration by \citet{Pecaut2013}. From the error in the {\it Gaia} 
parallax, the typical standard deviation in $\log(L/L_\odot)$ is estimated to 
be about 0.10\,dex, allowing for standard deviations of 0.1\,mag in the 
apparent magnitude, 0.1\,mag in visual extinction and 0.1\,mag in the 
bolometric correction in addition to the parallax error. 

Unfortunately, the effective temperature is not available for many stars.  It 
turns out that the derived luminosity is only slightly dependent on 
$T_{\rm eff}$, which is mainly used for deriving the bolometric correction.  
On the assumption that the selected stars are subdwarfs,  a good approximation to the luminosity for stars without
effective temperature measurements can be obtained simply by adopting $T_{\rm eff} 
\approx 30000$\,K for sdB and $T_{\rm eff} \approx 50000$\,K for sdO stars.  
The approximate luminosity obtained in this way is a good indicator of whether 
or not the star is a subdwarf.  Figure\,\ref{hrdiag} shows the theoretical H--R 
diagram for all {\em TESS} subdwarf candidates for which parallaxes are 
available.  

As can be seen, there are some stars well above the subdwarf sequence.  Stars 
with $\log(L/L_\odot)$ exceeding approximately 1 dex above the well-defined 
subdwarf sequence were eliminated from the sample.  There are also several 
stars cooler than about 15000\,K.  The effective temperatures for these stars 
are nearly all derived from photometric indices and are considered unreliable. 
On the assumption that they are all subdwarfs, and assigning $T_{\rm eff} =
30000$\,K to these stars, some were found to have luminosities consistent
with hot subdwarfs. As a  result, of the 1389 {\em TESS} subdwarf 
candidates examined, about 1250 lie within the hot subdwarf region.
These stars were retained for further study, while the others
were rejected.

\section{Known hot subdwarf pulsators}

Since the last compilation by \citet{Holdsworth2017}, several studies have 
identified more pulsating hot subdwarfs.  A list of these additional 
V361~Hya and V1093~Her stars is given in Table\,\ref{recent} in the Appendix.  
More recently, \citet{Groot2021} have identified new pulsating hot subdwarfs 
using {\em TESS} photometry.  These are also included in Table\,\ref{recent}.
Wherever possible, effective temperatures from the literature are included
in the tables.  Where no effective temperatures are available, the adopted 
effective temperatures described above were used to estimate the luminosity. 

\begin{figure}
\centering
\includegraphics[]{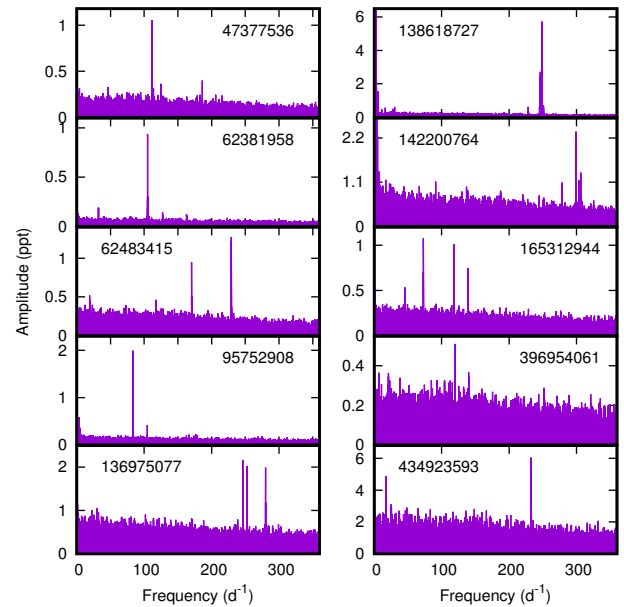}
\caption{Known pure p-mode pulsating hot subdwarfs observed by {\em TESS}
showing frequency peaks in the g-mode range.}
\label{pgmode}
\end{figure}

There are 26 know pure p-mode pulsators observed by {\em TESS} in the
compilation by \citet{Holdsworth2017}.  In ten of these stars significant 
frequency  peaks in the g-mode region were detected (Fig.\,\ref{pgmode}).  
These stars, listed in  Table\,\ref{sdpg}, clearly need to be re-classified as p+g-mode hybrids.
In this table, a list of previously unrecognised sdBV binary
candidates is also given.

\begin{table}
\setlength{\tabcolsep}{3pt}
\begin{center}
\caption{Stars in \citet{Holdsworth2017} classified as pure p-mode pulsators
but which have clear g-mode frequency peaks in the {\em TESS} photometry. 
In the second half of the table, known sdBV stars are listed in which
binary variation may be present. Effective temperatures are from \citet{Holdsworth2017}
and spectral types from the literature.}
\label{sdpg}
\resizebox{!}{3.6cm}{
\begin{tabular}{rlrrl}
\hline
\multicolumn{1}{r}{TIC}                      & 
\multicolumn{1}{l}{Name}                     & 
\multicolumn{1}{c}{$T_{\rm eff}$\,(kK)}      &
\multicolumn{1}{c}{$\log(L/L_\odot)$}        &
\multicolumn{1}{l}{Sp. Type}                 \\
\hline   
\\
\multicolumn{5}{l}{p-mode sdBV reclassified as hybrid pulsators:}        \\
  47377536 & V* UY Sex                   & $35.0 \pm 1.0$ & $1.59 \pm 0.07$ & sdO9VIIHe6     \\
  62381958 & GD 1053                     & $37.1 \pm 0.3$ & $1.52 \pm 0.08$ & sdO/Bsd+MSsdB  \\
  62483415 & PHL 252                     & $35.1 \pm 0.3$ & $1.46 \pm 0.07$ & sdO/BsdO/BsdO  \\
  95752908 & TYC 4890-19-1               & $34.2 \pm 0.5$ & $1.28 \pm 0.07$ & sdB            \\
 136975077 & KPD 2109+4401               & $31.8 \pm 0.6$ & $1.50 \pm 0.07$ & sdBV           \\
 138618727 & Feige 48                    & $29.5 \pm 0.2$ & $1.58 \pm 0.07$ & sdBV+dM        \\
 142200764 & HE 0230-4323                & $31.6 \pm 0.5$ & $1.62 \pm 0.08$ & sdB+dM         \\
 165312944 & PG 1219+534                 & $33.7 \pm 0.3$ & $1.46 \pm 0.07$ & sdBV           \\
 396954061 & GALEX J041550.2+015421      & $34.0 \pm 0.5$ & $1.45 \pm 0.07$ & sdBV           \\
 434923593 & [RWW88] CS 124636.2-631549  & $28.5 \pm 0.7$ & $1.20 \pm 0.07$ & sdBV+dM,       \\
\\
\multicolumn{5}{l}{New binary candidates:}                     \\
  80290366 &   CD-48 106                 & $25.2 \pm 1.2$ & $1.38 \pm 0.07$ & sdBV           \\
  80427831 &   EC 00404-4429             &                & $1.29 \pm 0.08$ & sdB+WD         \\
 117070953 &   TIC 117070953             &$170.0 \pm 2.0$ & $1.76 \pm 0.07$ &                \\
 115280751 &   SDSS J164214.21+425233.9  & $32.4 \pm 0.3$ & $1.93 \pm 0.10$ & sdBV+G0        \\
 137608661 &   TYC 4544-2658-1           &                & $1.41 \pm 0.07$ & sdB            \\
 138618727 &   Feige 48                  & $29.5 \pm 0.2$ & $1.58 \pm 0.07$ & sdBV+dM        \\
 142200764 &   HE 0230-4323              & $31.6 \pm 0.5$ & $1.62 \pm 0.08$ & sdB+dM         \\
 175402069 &   PG 1336-018               & $31.4 \pm 0.2$ & $1.32 \pm 0.07$ & sdBV+dM        \\
 220573709 &   BPS CS 31064-0017         & $52.0 \pm 3.0$ & $1.61 \pm 0.07$ & sdO            \\
 352480413 &   JL 82                     & $25.4 \pm 0.3$ & $1.38 \pm 0.07$ & sdB+dM         \\
 436579904 &   KUV 04421+1416            & $32.0 \pm 0.4$ & $1.24 \pm 0.07$ & sdBV+dM        \\
\\
\hline
\end{tabular}
}
\end{center}
\end{table}

The only meaningful subdivision of sdBV stars is the classification into 
p-mode and g-mode variables.  The restoring force in p-mode variables is the
pressure, while for g-mode variables it is buoyancy.  The 
subdivision is based on the expected frequency ranges for p- and g-modes derived 
from models. The theoretical pulsation period boundary between p- and g-modes is 
around 250\,s or a frequency of 340\,d$^{-1}$ \citep{Bloemen2014}.  This is
  the criterion used by \citet{Holdsworth2017}. In this paper we 
use  this definition of p- and g-mode variables. The shortest period 
that can be observed by {\em TESS} is 240\,s (2 min cadence), which means that 
only g-mode pulsators (V1096~Her stars) can be detected.

\section{New pulsational variables}

The {\em TESS} periodograms of hot subdwarfs have relatively low 
signal-to-noise ratios owing to the faintness of the stars.  It should also be noted 
that the {\em TESS} photometric bandpass is very wide and mostly sensitive in 
the near infrared \citep{Ricker2014}.  This is not well suited to detecting 
pulsations in very hot stars because the amplitudes are much lower in the near 
infrared than in the blue region of the spectrum.  

The periodograms of each star were visually inspected and the type of 
variability noted. Any star in which only one peak below 30\,d$^{-1}$ 
or a peak with one or more harmonics was detected is regarded as a binary. 
The dips in the light curves of detached eclipsing binaries manifest as 
harmonics in the periodogram.  The possibility exists that the 
variability is  due to pulsation modes with harmonics or even rotation, 
rather than binarity. The differences cannot be distinguished from the light 
curve alone.  On the other hand, if multiple peaks are seen that are not in 
a harmonic relation, then one may feel confident that pulsation is
involved.

In a few stars, a frequency peak with one or more harmonics is present in
addition to other unrelated frequency peaks.  These stars are classified as 
possible pulsating binary hot subdwarfs.

No significant light variation could be seen in 640 stars.
There are 24 stars that  are clearly main sequence stars with
erroneous parallaxes, and  12 stars with irregular light variations.
As many as 410 possible binaries were noted, but half of them have 
uncertain low amplitudes.  We classified 210 stars as definite
binary candidates,  8 of which were noted to have flares (presumably
originating in a cool companion).

Inspection of the periodograms of the {\em TESS} candidates within the
sdB region of the H--R diagram showed many of the variables have only a single 
significant peak in the periodogram.  These stars were removed from the 
sdBV candidate list on the assumption that the periodicity might be 
due to binary motion rather than pulsation, as described above.  After removing
a few other stars where the evidence for pulsation was considered too weak, 63 
previously unknown pulsating hot subdwarf candidates remain.  These include 38
stars in \citet{Groot2021}, which seemed to have been overlooked as sdBV
stars. They are listed in Table\,\ref{data1} and their periodograms shown
in Fig.\,\ref{sdb1}.  The remaining 25 stars are newly discovered sdBV
candidates (Table\,\ref{data2}). Periodograms of these stars are shown in 
Fig. \ref{sdb2}.

{\em TESS} images of every target star listed in Tables\,\ref{data1} and 
\ref{data2} were visually examined for the presence of background or 
neighbouring stars that could affect the signal in the {\em TESS} aperture. 
The target CCD pixel images were overlaid with scaled SDSS or 
{\em Digital Sky Survey} (DSS) images.  A visual assessment was made and the 
probable affect on the light curve designated by the code shown in 
Tables\,\ref{data1} and \ref{data2}.  The {\em TESS} pixel is $21 
\times 21$\,arcsec and in the worst case (code 3), intrusion of light from 
neighbouring stars of comparable brightness will affect the measured  light 
amplitude.

Whenever doubts have arisen about light intrusion into the {\em TESS}
aperture, the neighbouring stars were examined for light variability
when possible.  None of these neighbouring stars appeared to be a 
variable, but there were some cases where background stars occupied the 
same pixel as, or an adjoining pixel to the target star.  In these cases the field 
was examined using SDSS or DSS colour images to estimate the colours of
the neighbouring stars. In all cases the target appeared to be a blue 
star, while the neighbour(s) were distinctly cooler.  This makes it
less likely that the intruding star(s) are responsible for any of the 
pulsation patterns seen in the hot subdwarfs.

\section{Masses}

The luminosity derived from {\em Gaia} parallax provides an independent 
method of estimating the stellar mass. From the equations $L/L_\odot 
= (R/R_\odot)^2 (T_{\rm eff}/T_{\rm eff \odot})^4$ and
$g/g_\odot = M/M_\odot/(R/R_\odot)^2$ we have
\begin{align*}
\log(M/M_\odot) &= \log g + \log L/L_\odot - 4 \log T_{\rm eff} + 10.610.
\end{align*}
Since the $\log g$, $\log L/L_\odot$, and $\log T_{\rm eff}$ values are known for many
hot subdwarfs, it is possible to obtain individual stellar masses.  These masses might not be reliable because of the large errors in measuring these
three quantities.  The typical standard deviations are
$\sigma(\log T_{\rm eff}) \approx 0.010$, $\sigma(\log L/L_\odot) \approx 0.011$,
$\sigma(\log g) \approx 0.07$ which gives $\sigma(\log M/M_\odot) \approx 0.14$.
The expected error in mass is a substantial fraction of the mass
itself.  Clearly, obtaining individual stellar masses in this way is  not useful.  However, the average mass of a fairly large population will be
quite well determined and might prove interesting.

\begin{figure}
\centering
\includegraphics[]{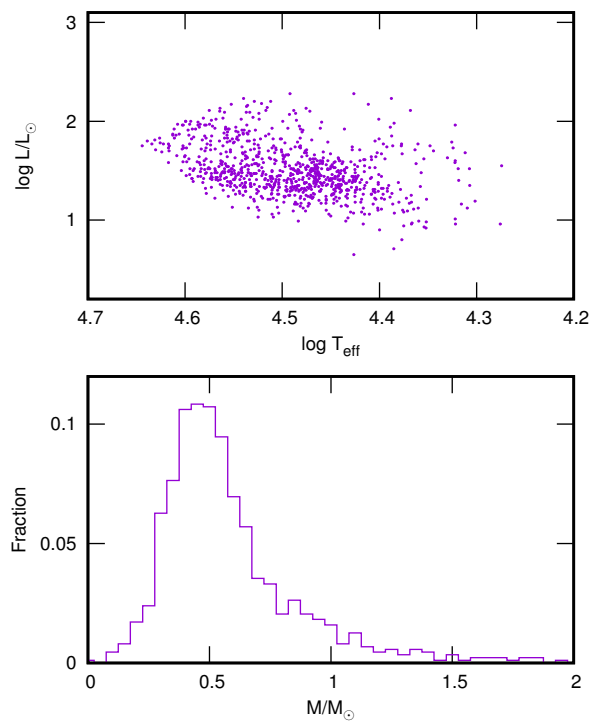}
\caption{Theoretical H--R diagram  of hot subdwarfs extracted 
from the catalogue of \citet{Geier2020} and located within the instability
region (top panel).  Bottom panel: Mass distribution of these stars estimated directly 
from the stellar parameters.}
\label{mhist}
\end{figure}

The catalogue of \citet{Geier2020} was used to extract 876 candidate subdwarfs
with known stellar parameters within the region of the p-mode and g-mode 
pulsators (top panel of Fig.\,\ref{mhist}).  The bottom panel of
Fig.\,\ref{mhist} shows the mass  distribution of these stars estimated from 
the above equation.  There is a long tail of high masses 
which could be a result of inclusion of main sequence stars in the 
\citet{Geier2020} catalogue. There are certainly a number of Be and other 
peculiar stars in the catalogue.  The peak of the distribution is at 
$M = 0.45$\,$M_\odot$.  If only stars in the range $0.25 < M/M_\odot < 1.0$ are 
retained, the mean is $M = 0.545 \pm 0.006$\,$M/M_\odot$ from 754 stars. 

Among  the 754 stars there are 45 g-mode pulsators with mean mass
$\overline{M/M_\odot} = 0.53 \pm 0.02$.  The 11 hybrids have 
$\overline{M/M_\odot} = 0.58 \pm0.04$, while the 27 p-mode pulsators have a 
mean mass $\overline{M/M_\odot} = 0.64 \pm 0.04$.  These values assume
that the measured values of effective temperature, luminosity, and surface
gravity are free of systematic errors.  Even if the actual values suffer
from systematic errors in the measured quantities, the relative values
should remain about the same. It is therefore safe to assume that p-mode 
pulsators have higher masses than g-mode pulsators.

Perhaps the most precise masses are those obtained from asteroseismology
(see \citealt{Fontaine2012}). From 20 asteroseismic masses of p-mode stars the
mean mass is $\overline{M/M_\odot} = 0.52 \pm 0.02$.  There are only three 
g-mode pulsators with asteroseismic masses, giving $\overline{M/M_\odot} = 
0.48 \pm 0.01$. Again, the p-mode pulsators have higher masses, but 
one can argue that a sample of three g-mode stars is too small for a 
meaningful determination. On the other hand, the \citet{Groot2013} dynamical 
and asteroseismic mass determination of PG1336-018, a sdBV in a binary system, 
seems to confirm the precision of asteroseismological models. The systematic 
difference between the asteroseismic masses and the masses derived here could 
be explained if the effective temperatures are increased by about 5\,\%.

In our list of newly discovered  pulsating hot subdwarfs, there are ten 
probable binaries, four of which have between 7 and 13 frequency peaks 
(Table\,\ref{sdpg}).  It might be possible to obtain both pulsation masses and 
dynamical masses from these stars.  This would allow an important verification 
of the pulsation masses.

\section{Location of the g-mode hot limit}

The models of \citet{Fontaine2003}, which can account for the high pulsation
frequencies in V386~Hya stars, require the observed modes to be of relatively 
high degree (spherical harmonic $l = 3$ or 4) because the more visible 
low-degree modes are only excited in stars cooler than the coolest known sdB 
stars.  Furthermore, the hot end of the g-mode instability strip turns out to 
be about $T_{\rm eff} \approx 24500$\,K, which is significantly cooler than 
observations.

The discrepancy between the observed and theoretical blue edges was resolved
by \citet{Jeffery2006b} using updated opacities together with enhanced Ni
and Fe in the driving zone.  They find the blue edge at around $T_{\rm eff}
\approx 28000$\,K for $l = 3$ if the Ni abundance is enhanced by a factor of 
20.  \citet{Hu2009} found that helium settling causes a shift in the theoretical blue 
edge of the g-mode instability domain to higher effective temperatures.  A 
blue edge at about $T_{\rm eff} \approx 29500$\,K is possible for $l = 3$ 
modes with Ni enhancement and helium settling.

\citet{Bloemen2014} point out that the abundance enhancements of Fe and Ni 
due to diffusion  were previously underestimated.  Their models predict g-mode 
pulsations at effective temperatures as high as $T_{\rm eff} = 33500$\,K which
is hotter than the observed hot limit for g-modes and well inside the p-mode 
region.


\begin{figure}
\centering
\includegraphics[]{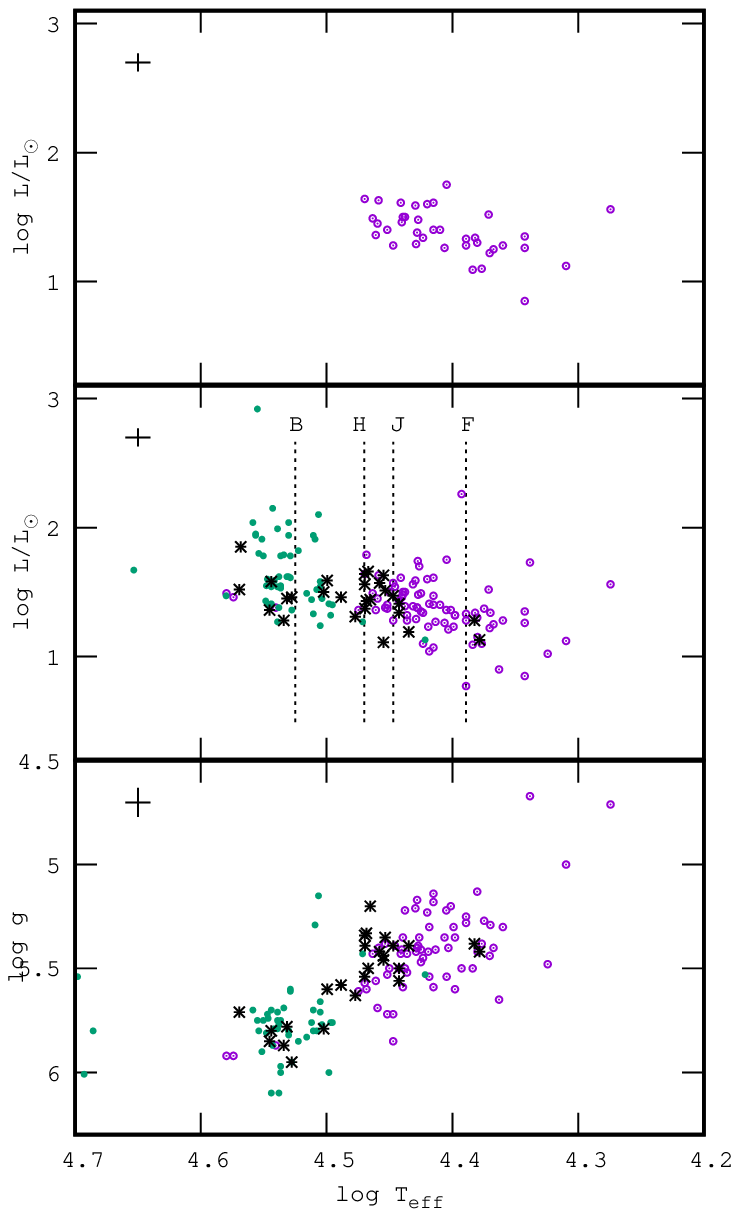}
\caption{Theoretical H--R diagram for {\em TESS} pulsating 
subdwarfs (top panel).  Middle panel: All pulsating subdwarfs, including the
{\em TESS} pulsators.  The p-mode pulsators are shown as green filled circles and 
the g-mode pulsators as  violet open circles.  The hybrid pulsators are shown 
as black asterisks. The four dotted lines show the predicted hot limit of 
\citet{Fontaine2003} (F), \citet{Jeffery2006b} (J), \citet{Hu2009} (H),
and \citet{Bloemen2014} (B).  Bottom panel: Gravity--effective temperature 
diagram for all pulsating subdwarfs. The cross at the top left of each panel shows
the approximate 1$\sigma$ error bars.}  
\label{hrvar}
\end{figure}

The top panel of Fig.\,\ref{hrvar} shows the location of the pulsating hot 
subdwarfs observed by {\em TESS} in the H--R diagram.  The middle panel shows 
all known pulsating hot subdwarfs as well as the four blue edges discussed 
above.  As already mentioned, many of the known p-mode {\em TESS} stars in the 
literature show distinct frequency peaks in the g-mode region as
well (Table\,\ref{sdpg} and Fig.\,\ref{pgmode}). These new  hybrid pulsators, 
as well as previously known hybrid pulsators, are shown in the middle panel of 
Fig.\,\ref{hrvar}.  They no longer form a tight sequence between the p- and 
g-mode pulsators, but are to be found right across the instability strip.
The bottom panel of Fig.\,\ref{hrvar} shows the same stars in the more
frequently used $\log g$ - $T_{\rm eff}$ diagram.

There is a simple reason for the lack of detection of g-mode frequency peaks in 
p-mode stars observed from the ground.  Because of daily gaps in
ground-based photometry, night-to-night changes in atmospheric conditions  and 
low amplitudes, low frequencies in the periodogram are contaminated by
the daily alias peaks and instrumental drift.  High frequency p-mode 
pulsations remain relatively unaffected.  On the other hand, the  almost 
continuous time sequences over many weeks obtained by {\em TESS} eliminates 
the aliasing problem and allows detection of low-frequency pulsations even at 
very low amplitudes. 

There are three very hot g-mode pulsators among the p-mode stars in
Fig.\,\ref{hrvar}: TIC\,442482669 (LS IV-14 116, \citealt{Ahmad2005}), 
TIC\,20420376 (UVO 0825+15, \citealt{Jeffery2017}), and TIC\,371813244 
(Feige 46, \citealt{Latour2019}). These pulsators may represent a different type of 
object \citep{Dorsch2020}.

\section{Conclusions}

A list of all known pulsating sdB stars was compiled by \citet{Holdsworth2017}.
This list was extended by a literature search, supplemented with the lists
from \citet{Groot2021} and presented in Table\,\ref{recent}.

By cross-matching the positions of stars observed by {\em TESS} with the lists 
of subdwarf candidates by \citet{Geier2020}, a total of 1398 stars with 
{\em TESS} light  curves was obtained.  Using the stellar luminosities derived 
from {\em Gaia} parallaxes, stars not likely to be hot subdwarfs were 
identified and removed from the sample. 

The periodograms of these stars were inspected and stars classified according 
to variability type. Pulsating hot subdwarfs are identified by the 
presence of multiple peaks in the periodogram.  Many other stars showing only 
one peak below 30\,d$^{-1}$, or one peak and its harmonics, are assumed to be 
binaries.

In this way, 63 pulsating hot subdwarfs were detected, 38 of which  are 
listed in \citet{Groot2021}, but not classified as pulsating variables.  The 
remaining 25 stars are previously unknown pulsating hot subdwarf candidates.  
This is a significant increase in the number of pulsating subdwarfs presented 
by the WG8 work. The new variables are presented in Tables\,\ref{data1} and 
\ref{data2}.  Their periodograms are shown in Figs.\,\ref{sdb1} and
\ref{sdb2}.

Stellar masses of a large sample of hot subdwarfs were  determined from 
$T_{\rm eff}$, $\log g,$ and  $\log L/L_\odot$.  Applying this method to a 
sample of p-mode and g-mode pulsators, it is found that masses of p-mode 
pulsators are 15\,\%\ higher than those of g-mode pulsators.  The same 
result is found using asteroseismic masses. 

While current pulsation models are able to explain the pulsations in these
stars, they are unable to fix the location of the g-mode blue edge.  The
models by \citet{Bloemen2014} allow g-mode pulsations to continue well 
past the observed blue edge, depending on the amount of iron-group metal
abundances in the driving zone. 

The {\em TESS} observations have detected many stars with g-modes previously
thought to be pure p-mode pulsators.  As a result, these newly classified
hybrid stars no longer form a tight group between the p- and g-mode stars. 
It seems likely that the previous observations could not detect the long 
periods of the g-mode stars, giving a misleading indication of the location
of the hybrid stars.  The situation is analogous to the lack of long-period
detections from the ground in $\delta$~Scuti stars.  Space observations
later showed that long periods (low frequencies) are present in almost all
$\delta$~Scuti variables \citep{Balona2014a}.

\begin{acknowledgements} LAB thanks the National Research Foundation of South Africa for financial 
support. This paper includes data collected by the {\it TESS} mission. Funding for the 
{\it TESS} mission is provided by the NASA Explorer Program. Funding for the 
{\it TESS} Asteroseismic Science Operations Centre is provided by the Danish 
National Research Foundation (Grant agreement no.: DNRF106), ESA PRODEX
(PEA 4000119301) and Stellar Astrophysics Centre (SAC) at Aarhus University. 

We thank the {\it TESS} and TASC/TASOC teams for their support of the present
work. This work has made use of data from the European Space Agency (ESA) mission
{\it Gaia} (\url{https://www.cosmos.esa.int/gaia}), processed by the {\it Gaia}
Data Processing and Analysis Consortium (DPAC,
\url{https://www.cosmos.esa.int/web/gaia/dpac/consortium}). Funding for the DPAC
has been provided by national institutions, in particular the institutions
participating in the {\it Gaia} Multilateral Agreement.

This research has made use of the SIMBAD database, operated at CDS, 
Strasbourg, France.  This research has made use of the VizieR catalogue access 
tool, CDS, Strasbourg, France (DOI: 10.26093/cds/vizier). The original 
description of the VizieR service was published in A\&AS 143, 23.

The data presented in this paper were obtained from the Mikulski Archive for 
Space Telescopes (MAST).  STScI is operated by the Association of Universities
for Research in Astronomy, Inc., under NASA contract NAS5-2655.

\end{acknowledgements}

\bibliographystyle{aa}
\bibliography{sdb}

\newpage

\begin{appendix}

\onecolumn


\section*{Appendix: Notes, tables and periodograms}

\begin{figure*}
\renewcommand\thefigure{A1}
\centering
\includegraphics[]{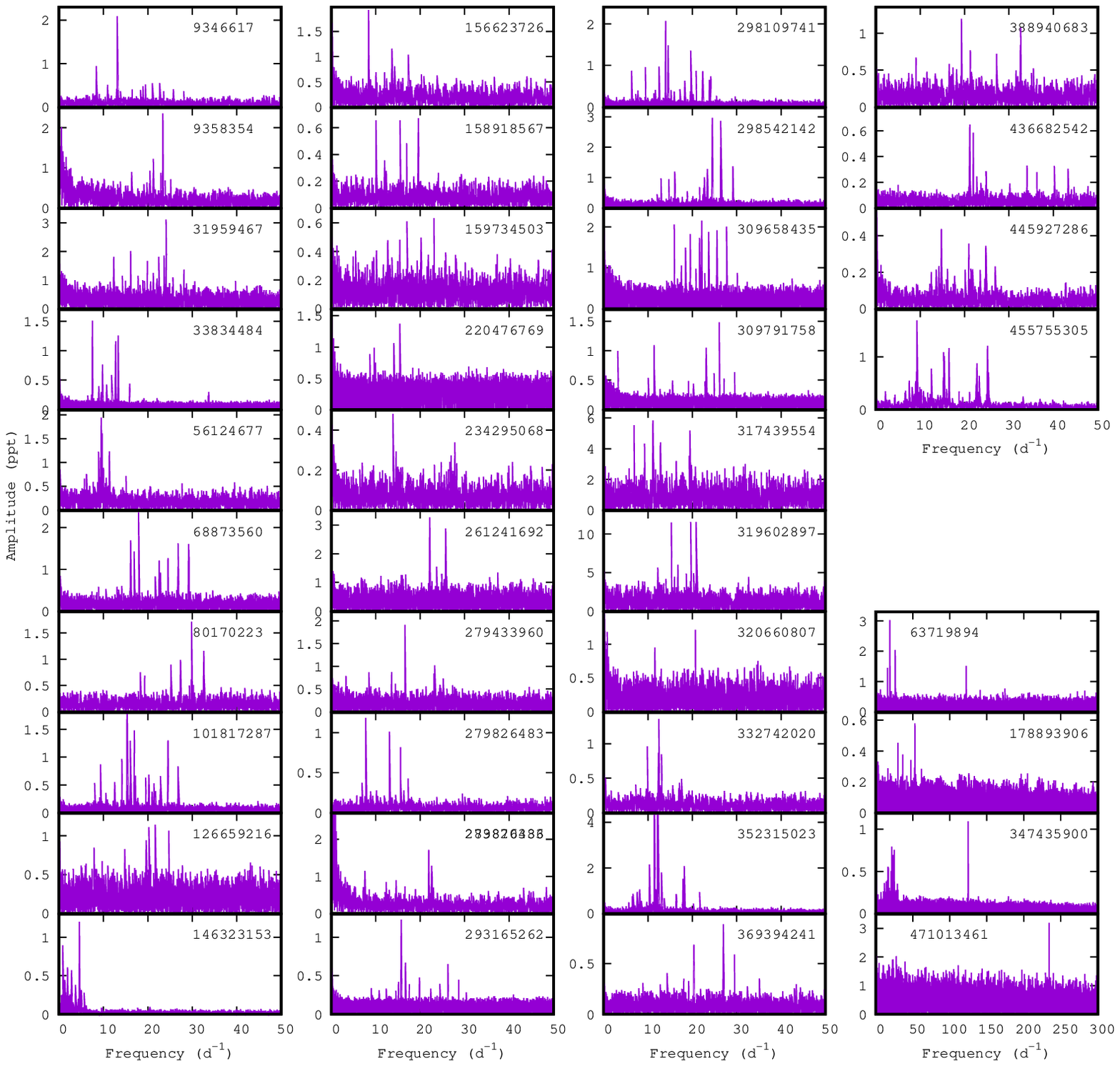}
\caption{Periodograms of unrecognised sdBV stars in \citet{Groot2021}.
The amplitude is in parts per thousand.}
\label{sdb1}
\end{figure*}

\begin{figure*}
\renewcommand\thefigure{A2}
\centering
\includegraphics[]{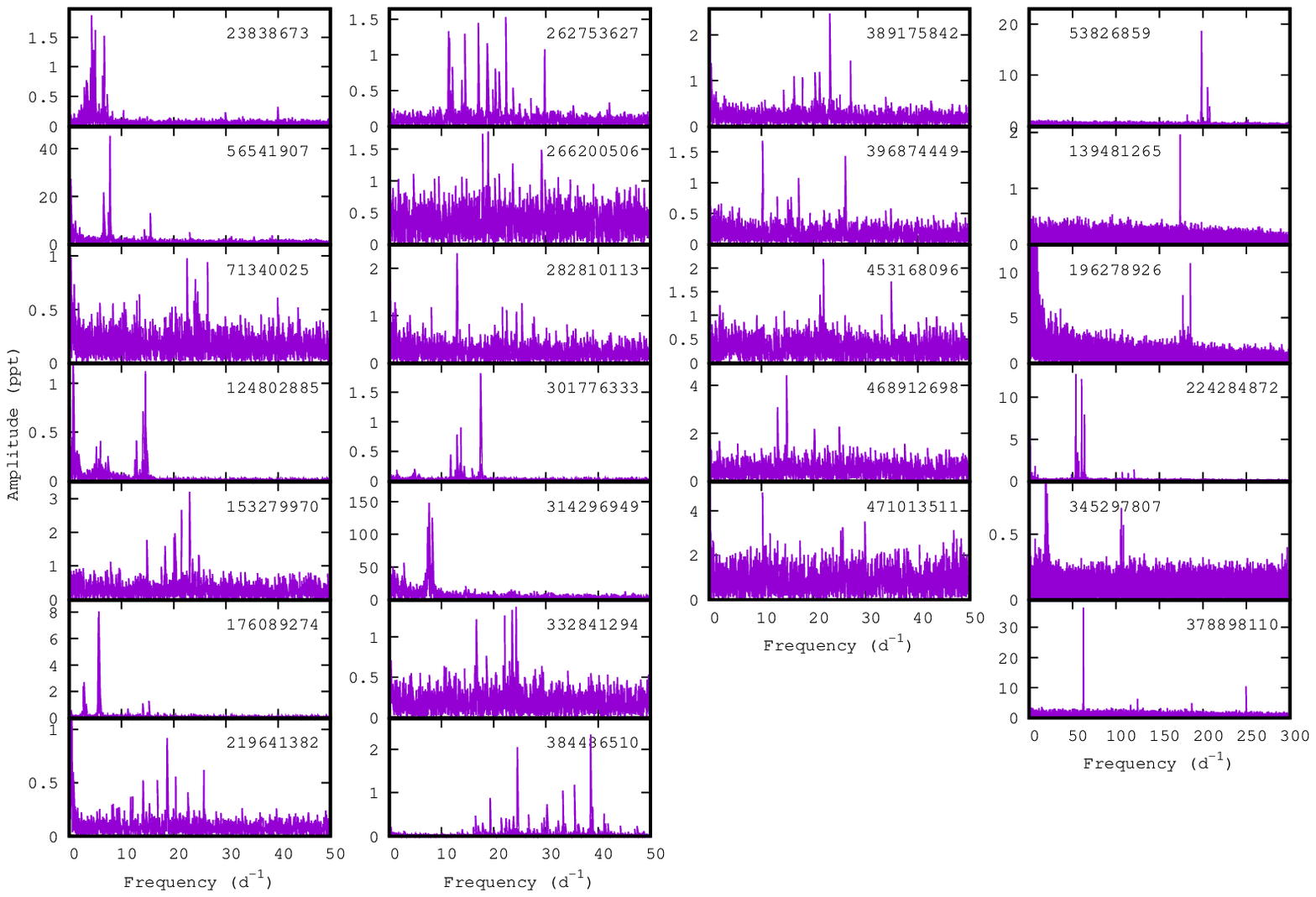}
\caption{Periodograms of newly discovered {\em TESS} sdBV stars.  The amplitude is in 
parts per thousand.}
\label{sdb2}
\end{figure*}

This appendix gives an updated list of known pulsating subdwarfs in
Table\,\ref{recent}, which continues the list presented in \citet{Holdsworth2017}.
Table\,\ref{data1} presents the {\em TESS} stars listed in \citet{Groot2021} that 
are unrecognised new pulsating variables.  Periodograms of these stars are 
shown in Fig.\,\ref{sdb1}.  Table\,\ref{data2} presents additional new 
{\em TESS} pulsating variables.  The periodograms of these stars are shown in 
Fig.\,\ref{sdb2}.


\begin{table*}
\setlength{\tabcolsep}{3pt}
\renewcommand\thetable{A1}
\begin{center}
\caption{Pulsating hot subdwarfs discovered since \citet{Holdsworth2017}. The 
pulsation type is either P (p-mode), G (g-mode), or H (both p- and g-modes present).  
Suspected binary systems are labelled B.  The effective temperature (kK) and
luminosity (from {\em Gaia} EDR3 parallaxes) are shown.  If no value of 
$T_{\rm eff}$ is available, the calculation of the luminosity assumes 
$T_{\rm eff} = 3.0$\,kK.  The surface gravity, $\log(g),$ and the log helium 
abundance $Y =$ n(He)/n(H) are from \citet{Geier2020}. The reference codes for 
pulsation (Pref) and for the effective temperatures (Tref) are listed in 
the last two columns. The second half of the table lists the {\em TESS}
sdBV stars discovered by \citet{Groot2021}.}
\label{recent}
\begin{tabular}{rllrrrrll}
\hline
\multicolumn{1}{r}{TIC}                      & 
\multicolumn{1}{l}{Name}                     & 
\multicolumn{1}{l}{Type}                     & 
\multicolumn{1}{c}{$T_{\rm eff}$\,(kK)}      &
\multicolumn{1}{c}{$\log(L/L_\odot)$}        &
\multicolumn{1}{c}{$\log(g)$}                &
\multicolumn{1}{c}{$\log(Y)$}                &
\multicolumn{1}{l}{Pref}                     & 
\multicolumn{1}{l}{Tref}                     \\
\hline   
         - &  DENIS J231105.0-013705    &  P  &  $35.35 \pm 0.60$ &  $1.43 \pm 0.16$  &                  &                  & 42  & 31,42              \\
   2621707 &  SDSS J135544.71-080354.3  &  G  &  $27.00 \pm 0.20$ &  $1.39 \pm 0.10$  &                  &                  & 20  & 13,20,22           \\
  13145616 &  CD-28 1974                &  G  &  $29.84 \pm 0.50$ &  $1.36 \pm 0.07$  &  $5.61 \pm 0.06$ & $-2.44 \pm 0.14$ & 40  & 12,14,40           \\
  20420376 &  UVO 0825+15               &  G  &  $37.98 \pm 0.40$ &  $1.49 \pm 0.07$  &  $5.92 \pm 0.10$ & $-0.62 \pm 0.08$ & 24  & 14,19,24,26        \\
  20448010 &  EC 11119-2405             &  G  &  $23.43 \pm 0.48$ &  $1.34 \pm 0.07$  &  $5.29 \pm 0.08$ & $-2.46 \pm 0.31$ & 43  & 14,43              \\
  21223262 &  SDSS J083603.98+155216.4  &  G  &  $26.24 \pm 0.50$ &  $1.23 \pm 0.09$  &  $5.42 \pm 0.07$ & $-2.45 \pm 0.17$ & 27  & 13,21,22,29,33     \\
  60257911 &  SDSS J082003.35+173914.2  &  P  &  $36.00 \pm 2.00$ &  $1.94 \pm 0.09$  &                  &                  & 27  & 27                 \\
  67584818 &  CD-33 417                 &  G  &  $25.06 \pm 0.50$ &  $1.23 \pm 0.07$  &  $5.30 \pm 0.20$ & $-2.80 \pm 0.00$ & 41  & 1,3,41             \\
  71013467 &  EC 04205-1328             &  G  &                   &  $2.32 \pm 0.07$  &                  &                  & 39  &                    \\
 117070953 &  TIC 117070953             &  H  & $170.00 \pm 2.00$ &  $1.76 \pm 0.07$  &                  &                  & 25  & 45                 \\
 137608661 &  TYC 4544-2658-1           & B+G &  $27.30 \pm 0.20$ &  $1.40 \pm 0.07$  &  $5.39 \pm 0.04$ & $-2.95 \pm 0.05$ & 46  & 46                 \\
 138707823 &  Ton S 135                 &  G  &  $25.00 \pm 0.30$ &  $1.32 \pm 0.07$  &  $5.60 \pm 0.20$ & $-2.30 \pm 0.20$ & 43  & 2,15,43            \\
 139723188 &  EC 21281-5010             &  P  &  $35.00 \pm 4.00$ &  $1.54 \pm 0.09$  &                  &                  & 35  & 35                 \\
 156618553 &  HW Vir                    &  H  &  $30.00 \pm 0.60$ &  $1.31 \pm 0.07$  &  $5.63 \pm 0.05$ & $ 0.00 \pm 0.00$ & 28  & 6,17               \\
 169285097 &  SB 815                    &  H  &  $27.20 \pm 0.55$ &  $1.25 \pm 0.05$  &  $5.39 \pm 0.10$ & $-2.94 \pm 0.01$ & 41  & 1,14,41            \\
 180205000 &  EC 11545-1459             &  P  &  $34.00 \pm 2.00$ &  $1.62 \pm 0.14$  &                  &                  & 35  & 35                 \\
 195189122 &  LB 378                    &  P  &  $35.24 \pm 0.60$ &  $1.60 \pm 0.13$  &  $5.72 \pm 0.11$ & $-1.92 \pm 0.19$ & 27  & 18,27,33,36        \\
 207440585 &  SDSS J161926.58+560558.6  &  P  &  $33.90 \pm 0.50$ &  $1.94 \pm 0.07$  &  $5.80 \pm 0.00$ & $-1.60 \pm 0.00$ &  9  & 10                 \\
 220573709 &  EC 03089-6421             &  P  &  $52.00 \pm 3.00$ &  $1.61 \pm 0.07$  &                  &                  & 35  & 35                 \\
 231846470 &  EC 01441-6605             &  P  &  $36.00 \pm 2.00$ &  $1.95 \pm 0.07$  &                  &                  & 35  & 35                 \\
 233689607 &  TIC 233689607             &  G  & $140.00 \pm 2.00$ &  $2.15 \pm 0.09$  &                  &                  & 25  &  45                \\
 260795163 &  EC 23073-6905             &  G  &  $27.00 \pm 0.50$ &  $1.56 \pm 0.07$  &                  &                  & 43  & 5,43               \\
 278659026 &  EC 21494-7018             &  G  &  $23.06 \pm 0.30$ &  $0.90 \pm 0.07$  &  $5.65 \pm 0.03$ & $-3.22 \pm 1.15$ & 32  & 14,15              \\
 335635628 &  PG 1315-123               &  H  &  $37.00 \pm 1.00$ &  $1.85 \pm 0.37$  &                  &                  & 34  & 34                 \\
 352444061 &  TIC 352444061             &  G  &                   &  $1.70 \pm 0.08$  &                  &                  & 25  &                    \\
 371813244 &  Feige 46                  &  G  &  $37.50 \pm 0.50$ &  $1.46 \pm 0.07$  &  $5.92 \pm 0.02$ & $-0.50 \pm 0.02$ & 37  & 3,33,56,57         \\
 382518318 &  GALEX J075010.4-644617    &  G  &                   &  $1.77 \pm 0.07$  &                  &                  & 39  &                    \\
 391825813 &  EC 10834-1301             &  P  &  $45.00 \pm 5.00$ &  $1.67 \pm 0.10$  &                  &                  & 35  & 35                 \\
 397064286 &  GALEX J050735.7+034815    &  G  &  $23.99 \pm 0.63$ &  $1.15 \pm 0.07$  &  $5.42 \pm 0.11$ & $-3.05 \pm 0.78$ & 44  & 14                 \\
 415339307 &  HS 0352+1019              &  G  &  $25.00 \pm 0.50$ &  $1.32 \pm 0.07$  &  $5.35 \pm 0.10$ & $-2.70 \pm 0.20$ & 43  &  7,26,43           \\
 418789164 &  HD 4539                   &  H  &  $24.12 \pm 0.20$ &  $1.28 \pm 0.07$  &  $5.38 \pm 0.05$ & $-2.42 \pm 0.20$ & 38  & 8,11,14,15,19,30   \\
 432223488 &  EC 15061-1442             &  P  &  $32.00 \pm 2.00$ &  $1.54 \pm 0.07$  &                  &                  & 35  & 35                 \\
 437043466 &  SDSS J083612.03+191755.9  &  H  &  $29.47 \pm 0.20$ &  $1.37 \pm 0.07$  &  $5.39 \pm 0.20$ & $-2.81 \pm 0.00$ & 23  & 14,21,23,26        \\
 446005482 &  TIC 446005482             &  G  & $135.00 \pm 2.00$ &  $2.76 \pm 0.10$  &                  &                  & 25  &  49                \\
 452718256 &  EC 11275-2504             &  P  &  $38.00 \pm 2.00$ &  $1.47 \pm 0.07$  &                  &                  & 35  & 35                 \\
 457168745 &  PG 0342+026               &  G  &  $26.00 \pm 1.00$ &  $1.07 \pm 0.07$  &  $5.59 \pm 0.12$ & $-2.69 \pm 0.10$ & 41  & 4,15,16            \\
\\
   9346617 &  KUV 09565+3632            &  G  &  $27.40 \pm 0.20$ &  $1.54 \pm 0.07$  &  $5.22 \pm 0.03$ & $-2.69 \pm 0.08$ & 47  & 21,26              \\
   9358354 &  GALEX J045213.2-091634    &  G  &                   &  $1.59 \pm 0.08$  &                  &                  & 47  &                    \\
  31959467 &  GALEX J191849.6-310441    &  G  &                   &  $1.43 \pm 0.07$  &                  &                  & 47  &                    \\
  33834484 &  [DI91] 1569               &  G  &  $24.10 \pm 0.50$ &  $1.34 \pm 0.07$  &                  &                  & 47  & 53                 \\
  56124677 &  SDSS J044246.86-071654.4  &  G  &  $22.00 \pm 0.50$ &  $1.26 \pm 0.08$  &                  &                  & 47  & 22                 \\
  63719894 &  FBS 0212+334              &  G  &                   &  $1.61 \pm 0.07$  &                  &                  & 47  &                    \\
  68873560 &  PG 1635+414               &  G  &  $26.85 \pm 0.42$ &  $1.29 \pm 0.07$  &  $5.42 \pm 0.06$ & $-2.75 \pm 0.24$ & 47  & 14                 \\
  80170223 &  BPS CS 22964-0098         &  G  &                   &  $1.53 \pm 0.07$  &                  &                  & 47  &                    \\
 101817287 &  EC 20106-5248             &  G  &  $24.50 \pm 0.50$ &  $1.33 \pm 0.09$  &  $5.25 \pm 0.12$ & $-2.77 \pm 0.10$ & 47  & 15,16              \\
 126659216 &  EC 20570-4308             &  G  &                   &  $1.52 \pm 0.08$  &                  &                  & 47  &                    \\
 146323153 &  GALEX J045547.2-203417    &  G  &                   &  $2.09 \pm 0.07$  &                  &                  & 47  &                    \\
 156623726 &  Ton 194                   &  G  &  $27.60 \pm 1.00$ &  $1.61 \pm 0.08$  &  $5.43 \pm 0.15$ & $-2.30 \pm 0.10$ & 47  & 50,58              \\
 158918567 &  TYC 3133-2416-1           &  G  &  $28.00 \pm 0.50$ &  $1.28 \pm 0.07$  &  $5.72 \pm 0.09$ & $-2.44 \pm 0.16$ & 47  & 26,56,57           \\
 159734503 &  BPS CS 29501-0054         &  G  &                   &  $1.58 \pm 0.08$  &                  &                  & 47  &                    \\
 178893906 &  EC 04284-2758             &  G  &                   &  $1.45 \pm 0.07$  &                  &                  & 47  &                    \\
 220476769 &  EC 05012-5641             &  G  &                   &  $1.71 \pm 0.09$  &                  &                  & 47  &                    \\
 234295068 &  EC 23483-6445             &  G  &                   &  $1.51 \pm 0.07$  &                  &                  & 47  &                    \\
 261241692 &  EC 19269-6231             &  G  &                   &  $1.62 \pm 0.08$  &                  &                  & 47  &                    \\
 279433960 &  BPS CS 22190-0004         &  G  &                   &  $1.50 \pm 0.08$   &                 &                  & 47  &                    \\
 279826483 &  US 719                    &  G  &                   &  $1.58 \pm 0.07$   &                 &                  & 47  &                    \\
\\                                                                                               
\hline
\end{tabular}
\end{center}
\end{table*}

\begin{table*}
\renewcommand\thetable{A1}
\begin{center}
\begin{tabular}{rllrrrrll}
\hline
\multicolumn{1}{r}{TIC}                      & 
\multicolumn{1}{l}{Name}                     & 
\multicolumn{1}{l}{Type}                     & 
\multicolumn{1}{c}{$T_{\rm eff}$\,(kK)}      &
\multicolumn{1}{c}{$\log(L/L_\odot)$}        &
\multicolumn{1}{c}{$\log(g)$}                &
\multicolumn{1}{c}{$\log(Y)$}                &
\multicolumn{1}{l}{Pref}                     & 
\multicolumn{1}{l}{Tref}                     \\
\hline   
 283870336 &  Ton 245                   &  G  &  $22.90 \pm 0.50$ &  $1.28 \pm 0.07$   & $5.30 \pm 0.15$ & $-2.50 \pm 0.10$ & 47  & 3,4                \\
 293165262 &  EC 04256-5912             &  G  &  $26.00 \pm 0.50$ &  $1.40 \pm 0.07$   &                 &                  & 47  & 53                 \\
 298109741 &  PG 1340+607               &  G  &                   &  $1.71 \pm 0.07$   &                 &                  & 47  &                    \\
 298542142 &  V* V1099 Her              &  G  &  $27.55 \pm 0.50$ &  $1.46 \pm 0.07$   & $5.41 \pm 0.03$ & $-2.70 \pm 0.07$ & 47  & 26                 \\
 309658435 &  EC 05155-6100             &  G  &  $25.70 \pm 0.40$ &  $1.40 \pm 0.08$   &                 &                  & 47  & 53                 \\
 309791758 &  EC 05201-6132             & B+G &                   &  $1.54 \pm 0.08$   &                 &                  & 47  &                    \\
 317439554 &  KPD 0716+0258             &  G  &  $26.30 \pm 1.00$ &  $1.60 \pm 0.08$   & $5.23 \pm 0.02$ & $-2.46 \pm 0.10$ & 47  & 26                 \\
 319602897 &  EC 19205-5916             &  G  &                   &  $1.51 \pm 0.08$   &                 &                  & 47  &                    \\
 320660807 &  GALEX J194855.8-583737    &  G  &                   &  $1.42 \pm 0.07$   &                 &                  & 47  &                    \\
 332742020 &  PG 0209-015               &  G  &  $23.50 \pm 0.50$ &  $1.52 \pm 0.08$   &                 &                  & 47  & 48                 \\
 347435900 &  Ton 209                   &  G  &  $29.50 \pm 1.00$ &  $1.64 \pm 0.07$   & $5.57 \pm 0.15$ & $-3.00 \pm 0.00$ & 47  & 3,4                \\
 352315023 &  JL 36                     &  G  &  $24.00 \pm 0.50$ &  $1.30 \pm 0.07$   & $5.13 \pm 0.16$ & $-2.67 \pm 0.20$ & 47  & 3,54,57            \\
 369394241 &  HE 0452-3654              &  G  &                   &  $1.48 \pm 0.07$   &                 &                  & 47  &                    \\
 388940683 &  PG 1230+052               &  G  &  $28.30 \pm 1.00$ &  $1.40 \pm 0.07$   & $5.72 \pm 0.15$ & $-3.00 \pm 0.10$ & 47  & 4                  \\
 436682542 &  GALEX J045437.4+140125    &  G  &                   &  $2.08 \pm 0.07$   &                 &                  & 47  &                    \\
 445927286 &  PG 0314+103               &  G  &                   &                    &                 &                  & 47  &                    \\
 455755305 &  FBS 2347+385              &  G  &  $23.80 \pm 0.50$ &  $1.10 \pm 0.07$   & $5.38 \pm 0.06$ & $-3.44 \pm 0.30$ & 47  & 14,17              \\
 471013461 &  EC 03530-4024             &  G  &                   &  $1.63 \pm 0.08$   &                 &                  & 47  &                    \\
\\                                                                                     
\hline
\\                                                                                 
\multicolumn{9}{l}{                                                                    
  1-\citet{Heber1984};  2-\citet{Heber1986};  3-\citet{Kilkenny1988};  4-\citet{Saffer1994};
  5-\citet{Kilkenny1995};
}\\
\multicolumn{9}{l}{                                                                    
  6-\citet{Hilditch1996}; 7-\citet{Edelmann2003}; 8-\citet{Sanchez-Blazquez2006};   9-\citet{Reed2007}; 
}\\
\multicolumn{9}{l}{                                                                    
 10-\citet{Ostensen2010c}; 11-\citet{Prugniel2011}; 12-\citet{Vennes2011}; 13-\citet{Girven2011}; 14-\citet{Nemeth2012};
}\\
\multicolumn{9}{l}{                                                                    
 15-\citet{Geier2013}; 16-\citet{Geier2013b}; 17-\citet{Kupfer2015}; 18-\citet{Kepler2015}; 19-\citet{Soubiran2016};
}\\
\multicolumn{9}{l}{                                                                    
 20-\citet{Bachulski2016}; 21-\citet{Luo2016}; 22-\citet{Perez-Fernandez2016}; 23-\citet{Baran2017};
}\\
\multicolumn{9}{l}{                                                                    
 24-\citet{Jeffery2017}; 25-\citet{Corsico2021}; 26-\citet{Lei2018}; 27-\citet{Reed2018b}; 28-\citet{Baran2018};
}\\
\multicolumn{9}{l}{                                                                    
 29-\citet{Reed2018}; 30-\citet{Arentsen2019}; 31-\citet{Anders2019};  32-\citet{Charpinet2019};
 33-\citet{Lei2019};
}\\
\multicolumn{9}{l}{                                                                    
 34-\citet{Reed2019}; 35-\citet{Kilkenny2019}; 36-\citet{Kepler2019};  37-\citet{Latour2019} ;
38-\citet{Silvotti2019};
}\\
\multicolumn{9}{l}{                                                                    
 39-\citet{Pelisoli2020}; 40-\citet{Reed2020}; 41-\citet{Sahoo2020}; 42-\citet{Ostensen2020};
 43-\citet{Uzundag2021} ;
}\\
\multicolumn{9}{l}{                                                                    
 44-\citet{Schaffenroth2018}; 45-\citet{Holberg1998};
46-\citet{Silvotti2021}; 47-\citet{Groot2021}.
}\\
\\
\hline
\end{tabular}
\end{center}
\end{table*}

\begin{table*}
\renewcommand\thetable{A2}
\begin{center}
\caption{List of 38 newly discovered pulsating sdB stars  not marked 
as pulsators in the  \citet{Groot2021} catalogue.  The columns are the
same as in Table\,\ref{recent}.  An asterisk refers to a note for the star.  
The column marked F gives the following flags: 1 - an isolated 
target or target star with neighbours that have no influence on the target 
fluxes; 2 - a crowded field where the target flux is not substantially 
polluted by light from neighbouring stars; 3 - a star with bright neighbours 
that might substantially affect the amplitudes of the subdwarf light 
variation. The last column is the spectral type from 
\citet{Groot2021}.}
 
\label{data1}
\resizebox{!}{7cm}{
\begin{tabular}{rllrlrrrrl}
\hline
\multicolumn{1}{r}{TIC}                      & 
\multicolumn{1}{l}{Name}                     & 
\multicolumn{1}{l}{Type}                     & 
\multicolumn{1}{c}{$T_{\rm eff}$\,(kK)}      &
\multicolumn{1}{l}{Tref}                     & 
\multicolumn{1}{c}{$\log(L/L\odot)$}         &
\multicolumn{1}{c}{$\log(g)$}                &
\multicolumn{1}{c}{$\log(Y)$}                &
\multicolumn{1}{c}{F}                        &
\multicolumn{1}{l}{Type}                 \\
\hline
  21343832* &  HS 0740+3734                &  G    &  $20.40 \pm 0.90$ &  7        & $1.12 \pm 0.08$ & $5.00 \pm 0.20$ & $-2.10 \pm 0.40$ &  1 &  sdB      \\  
  40050637~ &  KUV 16256+4034              &  G    &  $23.46 \pm 0.50$ & 14,17,26  & $1.22 \pm 0.07$ & $5.44 \pm 0.07$ & $-3.08 \pm 0.29$ &  2 &  sdB      \\  
  63208546* &  KIC 8754603                 & B+G   &                   & 56        & $1.25 \pm 0.07$ & $6.09 \pm 0.02$ & $-2.43 \pm 0.13$ &  2 &  sdB+?    \\  
  70963660~ &  EC 04178-1734               &  G    &                   &           & $1.40 \pm 0.07$ &                 &                  &  3 &  sdB      \\  
  88484868~ &  FBS 0658+627                &  G    &  $28.75 \pm 0.37$ &  14       & $1.63 \pm 0.07$ & $5.40 \pm 0.07$ & $-2.76 \pm 0.26$ &  3 &  sdB      \\  
 118298029~ &  GALEX J021618.9+275900      &  G    &  $25.48 \pm 0.45$ &  14       & $1.26 \pm 0.07$ & $5.35 \pm 0.05$ & $-2.81 \pm 0.23$ &  1 &  sdB      \\  
 137502282~ &  TIC 137502282               &  G    &                   &           & $1.39 \pm 0.07$ &                 &                  &  2 &  B1,sdB   \\  
 151641733* &  GALEX J110743.0-373158      &  G    &                   &           & $2.29 \pm 0.07$ &                 &                  &  2 &  sdB+MS   \\  
 154510451~ &  FBS 1224+780                &  G    &                   &           & $1.50 \pm 0.07$ &                 &                  &  1 &  sdB      \\  
 199732600~ &  HE 0127-43250               &  G    &                   &           & $1.57 \pm 0.08$ &                 &                  &  1 &  sdB      \\  
 201251043* &  TIC 201251043               &  G    &                   &           & $1.54 \pm 0.07$ &                 &                  &  1 &  sdB      \\  
 202354658~ &  PG 1544+601                 &  G    &  $29.06 \pm 0.50$ &  14       & $1.49 \pm 0.08$ & $5.43 \pm 0.05$ & $-3.44 \pm 0.60$ &  1 &  sdB      \\  
 229706981~ &  TIC 229706981               &  G    &                   &           & $1.58 \pm 0.11$ &                 &                  &  3 &  sdB      \\  
 232521983~ &  TYC 4407-187-1              &  G    &                   &           & $1.55 \pm 0.07$ &                 &                  &  1 &  sdB      \\  
 233211303~ &  HS 1747+6924                &  G    &  $27.50 \pm 1.20$ &   7       & $1.50 \pm 0.08$ & $5.35 \pm 0.20$ & $-2.90 \pm 0.30$ &  1 &  sdB      \\  
 240109525~ &  GALEX J064347.7+320147      &  G    &  $26.88 \pm 0.50$ &  26       & $1.59 \pm 0.08$ & $5.21 \pm 0.01$ & $-2.70 \pm 0.05$ &  3 &  sdB      \\  
 240783347~ &  TIC 240783347               &  G    &                   &           & $1.36 \pm 0.07$ &                 &                  &  1 &  sdB      \\  
 269766236~ &  TIC 269766236               &  G    &                   &           & $1.53 \pm 0.07$ &                 &                  &  1 &  sdB      \\  
 273218137~ &  TIC 273218137               &  G    &                   &           & $1.55 \pm 0.07$ &                 &                  &  3 &  sdB      \\  
 274623605~ &  PG 1700+48605               &  G    &  $26.01 \pm 1.11$ & 14        & $1.61 \pm 0.07$ & $5.18 \pm 0.09$ & $-2.58 \pm 0.33$ &  3 &  sdB      \\  
 279373920* &  TIC 279373920               & B+G   &                   &           & $1.51 \pm 0.07$ &                 &                  &  1 &  sdB      \\
 308179612~ &  EC 13080-1508               &  G    &                   &           & $1.41 \pm 0.07$ &                 &                  &  1 &  sdB      \\
 311432346* &  BD+29 3070                  & B+G   &  $25.38 \pm 0.99$ & 52        & $1.75 \pm 0.07$ & $5.54 \pm 0.18$ & $-2.63 \pm 0.00$ &  1 &  sdB F     \\
 311898870* &  FBS 2202+436                & B+G   &                   &           & $1.41 \pm 0.07$ &                 &                  &  1 &  sdB      \\
 331553315~ &  FBS 0430+772                &  G    &  $26.80 \pm 0.70$ &  7        & $1.38 \pm 0.07$ & $5.40 \pm 0.10$ & $-3.00 \pm 0.20$ &  3 &  sdB      \\
 352412700~ &  KUV 20417+7604              &  G    &                   &           & $1.55 \pm 0.07$ &                 &                  &  2 &  sdB      \\
 365496228~ &  TIC 365496228               &  G    &  $22.00 \pm 1.00$ & 13        & $0.85 \pm 0.07$ &                 &                  &  3 &  sdB      \\
 367003034~ &  80 GALEX J223336.8+741254   &  G    &                   &           & $1.57 \pm 0.07$ &                 &                  &  2 &  sdB      \\
 380641758* &  BD+10 2357                  &  G    &                   &           & $2.06 \pm 0.07$ &                 &                  &  1 &  sdO+A    \\
 387107334~ &  BPS CS 22959-0140           &  G    &                   &           & $1.47 \pm 0.08$ &                 &                  &  2 &  sdB      \\
 405799245~ &  PG 0926+06545               &  G    &  $26.50 \pm 0.68$ & 14,26     & $1.34 \pm 0.07$ & $5.45 \pm 0.08$ & $-2.79 \pm 0.46$ &  2 &  sdB      \\
 408552372~ &  EC 12578-2107               &  G    &                   &           & $1.53 \pm 0.07$ &                 &                  &  1 &  sdB      \\ 
 421895532* &  JL 111                      & B+G   &                   &           & $1.30 \pm 0.07$ &                 &                  &  1 &  sdB      \\
 424383364~ &  HS 1831+7647                &  G    &  $23.30 \pm 0.60$ &  7        & $1.25 \pm 0.07$ & $5.40 \pm 0.10$ & $-3.00 \pm 0.20$ &  3 &  sdB      \\
 437746793~ &  PG 0011+283                 &  G    &  $24.20 \pm 0.50$ & 33,56,57  & $1.09 \pm 0.07$ & $5.50 \pm 0.03$ & $-4.05 \pm 0.28$ &  1 &  sdB      \\
 457225725* &  TIC 457225725               &  G    &                   &           & $1.24 \pm 0.07$ &                 &                  &  1 &  sdB      \\
 466277784* &  EC 20182-6534               & B+G   &                   &           & $1.46 \pm 0.07$ &                 &                  &  1 &  sdB+WD   \\
 468980287~ &  GALEX J083412.3+071211      &  G    &  $28.90 \pm 0.50$ & 13,14,26  & $1.36 \pm 0.08$ & $5.56 \pm 0.12$ & $-2.62 \pm 0.00$ &  2 &  sdB      \\
\\
\hline
\\                                                                                                                                             
\multicolumn{10}{l}{48-\citet{Moehler1990}; 49-\citet{Ramspeck2001}; 
50-\citet{Maxted2001}; 51-\citet{Arkhipova2002}; 52-\citet{Vos2013};}\\
\multicolumn{10}{l}{53-\citet{MoniBidin2017};  54-\citet{Geier2017b}; 
 55-\citet{Bai2018};  56-\citet{Luo2019}; 57-\citet{Geier2020}; 
58-\citet{Tian2020};}\\
\\
\hline
\end{tabular}
}
\end{center}
\end{table*}

\section*{Notes to Table\,A2}

{\bf TIC\,21343832}. A flare at BJD\,2458844.77 seems to be present.
\\
{\bf TIC\,63208546}. A peak at 3.375\,d$^{-1}$ with a very strong harmonic
suggests a double-wave light curve with period $P=0.296$\,d from orbital
motion.
\\
{\bf TIC\,151641733}. \citet{Pelisoli2020} attributed the peak at
1.79\,d$^{-1}$ as being due to rotation of the secondary.  However, there are several peaks in the range 22.7--29.7\,d$^{-1}$ with S/N between 6 and 7
which are likely g-mode pulsations.
\\
{\bf TIC\,201251043, 457225725}. These stars were analysed by \citet{Sahoo2020b} using {\em TESS} full-frame images sampled with a cadence of 30\,min.  Although they are listed as pulsating sdB stars, the long cadence prohibits the detection of frequencies higher than about 24\,d$^{-1}$.  They all have frequency peaks exceeding this limit.
\\
{\bf TIC\,279373920}. A high-amplitude peak at 2.126\,d$^{-1}$ and its
harmonic indicates binary motion with $P = 0.468$\,d.
\\
{\bf TIC\,311432346}. The light curve shows clear sinusoidal variations with
a period $P = 1.495$\,d, which is perhaps orbital.
\\
{\bf TIC\,311898870}. The main frequency peak and a marginal harmonic
suggest orbital motion with $P = 0.363$\,d.
\\
{\bf TIC\,380641758}.  The pulsation frequencies in this star do not
correspond to those generally seen in sdB or sdO stars.  Typical sdO
pulsations are in the p-mode region, which is higher than the {\em TESS}
Nyquist frequency.  Perhaps they originate in the A-type companion.
\\
{\bf TIC\,421895532}. A peak at 1.314\,d$^{-1}$ and its harmonic indicates
binary motion with $P = 0.761$\,d.  The harmonic to the main peak at 
6.896\,d$^{-1}$ is also present.  Perhaps 0.761\,d represents rotation
and 0.145\,d the orbital motion.
\\
{\bf TIC\,466277784}. The strongest peak at 1.667\,d$^{-1}$ and its harmonic
suggests binary motion with period $P = 0.599$\,d.

\begin{table*}
\renewcommand\thetable{A3}
\begin{center}
\caption{List of 25 newly discovered {\em TESS} pulsating sdB candidates that  
are not listed in \citet{Groot2021}. The columns are the same as in 
Table\,\ref{data1}. The last column is the spectral type from the literature.}
\label{data2}
\resizebox{!}{4.5cm}{
\begin{tabular}{rllrlrrrrl}
\hline
\multicolumn{1}{r}{TIC}                      & 
\multicolumn{1}{l}{Name}                     & 
\multicolumn{1}{l}{Type}                     & 
\multicolumn{1}{c}{$T_{\rm eff}$\,(kK)}      &
\multicolumn{1}{l}{Tref}                     & 
\multicolumn{1}{c}{$\log(L/L\odot)$}         &
\multicolumn{1}{c}{$\log(g)$}                &
\multicolumn{1}{c}{$\log(Y)$}                &
\multicolumn{1}{c}{F}                        &
\multicolumn{1}{l}{Type}                 \\
\hline
  23838673* &  PG 1400+389             &  G    &  $18.80 \pm 1.00$ & 49          &  $1.56 \pm 0.07$  &  $4.71 \pm 0.10$ & $-0.71 \pm 0.00$ &  1 &  sd:O    \\  
  53826859~ &  TIC 53826859            &  G    &                   &             &  $2.02 \pm 0.09$  &                  &                  &  3 &          \\  
  56541907* &  TIC 56541907            & B+G   &                   &             &  $1.55 \pm 0.07$  &                  &                  &  2 &          \\  
  71340025~ &  GALEX J101850.0-333150  &  G    &                   &             &  $1.56 \pm 0.07$  &                  &                  &  3 &  sdO/B   \\
 124802885* &  HD 154137               &  G    &                   &             &  $1.22 \pm 0.07$  &                  &                  &  3 &          \\  
 139481265~ &  TIC 139481265           &  G    &                   &             &  $1.22 \pm 0.07$  &                  &                  &  3 &          \\
 153279970~ &  TIC 153279970           &  G    &                   &             &  $1.47 \pm 0.07$  &                  &                  &  3 &          \\  
 176089274* &  TIC 176089274           & B+G   &                   &             &  $2.08 \pm 0.09$  &                  &                  &  3 &          \\  
 196278926~ &  TIC 196278926           &  G    &                   &             &  $1.31 \pm 0.07$  &                  &                  &  3 &          \\  
 219641382* &  EC 11383-2238           & B+G   &                   &             &  $1.69 \pm 0.07$  &                  &                  &  2 &  sdB+MS  \\  
 224284872* &  SB 825                  & B+G   &                   &             &  $1.60 \pm 0.07$  &                  &                  &  1 &  A3      \\  
 262753627* &  TIC 262753627           &  G    &  $28.80 \pm 0.50$ & 26,56,57    &  $1.45 \pm 0.07$  &  $5.69 \pm 0.02$ & $-2.64 \pm 0.03$ &  3 &          \\  
 266200506~ &  GALEX J075215.1+004711  &  G    &                   &             &  $1.48 \pm 0.08$  &                  &                  &  3 &  sdB     \\  
 282810113* &  TIC 282810113           &  G    &                   &             &  $1.73 \pm 0.07$  &                  &                  &  3 &          \\
 301776333~ &  HD 289333               &  G    &                   &             &  $2.38 \pm 0.07$  &                  &                  &  3 &   B5     \\
 314296949* &  TIC 314296949           &  G    &                   &             &  $1.32 \pm 0.07$  &                  &                  &  3 &          \\
 332841294~ &  TIC 332841294           &  G    &                   &             &  $1.48 \pm 0.07$  &                  &                  &  3 &          \\
 345297807~ &  PG 1444+236             &  G    &  $22.00 \pm 0.50$ & 51          &  $1.35 \pm 0.07$  &                  &                  &  1 &   sd     \\
 378898110~ &  TIC 378898110           &  G    &                   &             &  $0.59 \pm 0.07$  &                  &                  &  2 &          \\
 384486510~ &  CPD-79 906              &  G    &                   &             &  $2.22 \pm 0.07$  &                  &                  &  1 &   B8     \\
 389175842~ &  PG 0940+068             &  G    &  $26.74 \pm 1.00$ & 33,56       &  $1.48 \pm 0.08$  &  $5.39 \pm 0.04$ & $-2.93 \pm 0.07$ &  3 &   sdB+WD,DA2: \\
 396874449~ &  PG 1111-077             &  G    &                   &             &  $1.33 \pm 0.07$  &                  &                  &  1 &   sdOB   \\
 453168096~ &  TIC 453168096           &  G    &                   &             & $-0.19 \pm 0.13$  &                  &                  &  2 &          \\
 468912698~ &  GALEX J075257.2+055911  &  G    &                   &             &  $1.46 \pm 0.08$  &                  &                  &  3 &   sdB    \\
 471013511* &  BPS CS 22879-0149       &  G    &  $24.50 \pm 0.50$ & 19          &  $1.28 \pm 0.08$  &                  &                  &  1 &   sdB    \\
\\
\hline                                                                                          
\\
\end{tabular}
}
\end{center}
\end{table*}

\section*{Notes to Table\,A3}

{\bf TIC\,23838673}. A forest of frequencies below 10\,d$^{-1}$ indicate 
a main sequence $\beta Cep$. There is a cut-off frequency limit for sdBV
pulsation mode excitation (about 5\,d$^{-1}$) below which we do not see any
pulsation due to energy dispersion in the driving zone. However, in this star 
there are some higher frequency modes of low amplitudes visible in the 
periodogram, and we left it for further investigations.
\\
{\bf TIC\, 56541907}. The presence of a strong peak at 7.770\,d$^{-1}$ and
its harmonic suggest a binary with period $P = 0.129$\,d.
\\
{\bf TIC\,124802885 (HD 154137)} and {\bf TIC\,301776333 (HD 289333)}.
 While both are classified as  normal main sequence stars, their
luminosities are well within the hot subdwarf range.  They are not listed as 
subdwarf candidates in the literature. It is possible that the {\em Gaia} parallaxes
are in error.
\\
{\bf TIC\,176089274}.  The peak at $5.608$\,d$^{-1}$ has a weak harmonic
suggesting orbital motion with $P=0.178$\,d.
\\
{\bf TIC\,219641382}. The light curve shows a clear sinusoidal variation
with $P = 2.7$\,d.
\\
{\bf TIC\,224284872}. This high-lattitude faint star is listed as a candidate 
hot subdwarf in \citet{Geier2019} and also in \citet{Gontcharov2011}.  The 
luminosity is certainly compatible with a subdwarf (no reliable effective
temperature has been measured).  However, it was   given an A0
classification by \citet{Slettebak1971} and listed as a high-frequency
A-type main sequence pulsator by \citet{Holdsworth2014}.  On the basis of
the low luminosity it is a likely V1093~Her star with a rich frequency
spectrum in the range 50--130\,d$^{-1}$.  There are several interesting 
relationships between frequencies that suggest tidal interaction in a binary 
system.  The presence of a peak at 0.734\,d$^{-1}$ and its high-amplitude 
harmonic can be seen as a double-wave variation in the light curve.  This 
could be a result of binary motion with period $P = 1.362$\,d.  The star might 
turn out to be a main sequence star with incorrect parallax, but it certainly 
deserves attention.
\\
{\bf TIC\,262753627}. The star was also analysed by \citet{Sahoo2020b} using 
{\em TESS} full-frame images sampled with a cadence of 30\,min. 
It is not listed in \citet{Groot2021}. 
\\
{\bf TIC\,282810113}.  While the single peak at 13.001\,d$^{-1}$ is present, 
there are four other peaks with  4 < S/N < 5. However, confirmation of these 
frequencies is clearly required.
\\
{\bf TIC\,314296949} High amplitude frequencies in the range 5--10\,d$^{-1}$ are
unusual in an sdBV star.
\\
{\bf TIC\,471013511} The only significant peak at 10.25\,d$^{-1}$ may 
be due to  low-amplitude binary variations, either rotation or pulsation. While it is
listed as a cataclysmic binary or related object by \citet{Ritter2004},
it lies well within the sdB instability strip among the g-mode pulsators.



\end{appendix}

\end{document}